# Will you donate money to a chatbot? The effect of chatbot anthropomorphic features and persuasion strategies on willingness to donate

Ekaterina Novozhilova

Boston University, ekaterin@bu.edu

Jiacheng Huang

University of Minnesota

Le He

Boston University

Ziling Li

Boston University

James Cummings

Boston University

This work investigates the causal mechanism behind the effect of chatbot personification and persuasion strategies on users' perceptions and donation likelihood. In a 2 (personified vs. non-personified chatbot) x 2 (emotional vs. logical persuasion strategy) between-subjects experiment ($N = 76$), participants engaged with a chatbot that represented a non-profit charitable organization. The results suggest that interaction with a personified chatbot evokes perceived anthropomorphism, however, does not elicit greater willingness to donate. In fact, we found that commonly used anthropomorphic features, like name and narrative, led to negative attitudes toward an AI agent in the donation context. Our results showcase a preference for non-personified chatbots paired with logical persuasion appeal, emphasizing the significance of consistency in chatbot interaction, mirroring human-human engagement. We discuss the importance of moving from exploring the common scenario of chatbot with machine identity vs. chatbot with human identity in light of the recent regulations of AI systems.

• Human-centered computing~Human computer interaction (HCI)~HCI design and evaluation methods~User studies

**Additional Keywords and Phrases:** User experience design, Artifact or system, Empirical study that tells us about how people use a system, Quantitative methods

## 1 INTRODUCTION

Chatbots have been used in organizational settings for various objectives, such as customer support [13], recommendations [31], and social listening [47]. Following these industry practices, researchers have been exploring whether chatbots can improve customer support performance [38] and the main factors influencing customer experience with chatbots [32]. Human-chatbot interaction has also been extensively studied in marketing and sales, where chatbots are used to spread brand awareness, influence marketing, and build customer relationships to prompt future purchases [24]. Yet among the unexplored domains where chatbots can be successfully utilized are NGOs, fundraising organizations, and charities. Non-profit institutions are famously understaffed, and following the example of commercial enterprises, this problem can be solved by deploying chatbot systems.

While chatbots have been fulfilling multiple organizational functions, the understanding of human-chatbot interaction leaves much room for further exploration. One branch of the current chatbot interaction research comes from the investigation of human-like features and their impact on people's attitudes and behavior. In line with the dominant computer are social actors (CASA) paradigm [30], researchers have been investigating whether adding human-like qualities to AI systems leads to more human-like treatment of interfaces and, in turn, yield certain desirable outcomes that would be found when interacting with other humans. This tenet of the CASA paradigm was supported in the chatbot interaction setting. For instance, Go and Sundar [14] found that anthropomorphic cues–image of a human compared to a bubble figure–and human identity cues–a chatbot agent as a human versus a machine–lead to more favorable attitudes toward chatbots. As the recent regulations oblige companies to disclose bots' identities [9], we decided to manipulate the presence of anthropomorphic cues rather than human identity cues in this work. Based on the previous research, we investigate the effect of a chatbot that has a name and background information (personified chatbot) versus the one that does not have any of those features (non-personified chatbot).

Previous research has also highlighted the importance of consistency in computer-agent interactions. For instance, the asynchronicity of verbal and non-verbal cues has been shown to lower trust and cause negative attitudes toward an AI agent [15]. Likewise, inconsistency in tasks and agents can lead to similar negative outcomes. On the one hand, the research suggests that a more affective approach in human-chatbot interaction, such as the use of emotional appeal for persuasion [44] or framing the chatbot as an empathetic agent [40] can yield better results in the donation context. On the other hand, scholars argue that AI systems are perceived as less functionally capable in tasks that involve human competencies such as empathy or affect [5], [27].Thus, chatbots that use logic and reasoning in persuasion might work better in the donation context [44].

The present study aims to investigate whether anthropomorphic cues, such as name and background information, of a chatbot will elicit greater willingness to donate. We also explore whether this effect will be mediated by perceived anthropomorphism and chatbot perceptions. Finally, we propose to explore whether the persuasion strategy will have an interaction effect with chatbot personification.

## 2 LITERATURE REVIEW

### 2.1 Perceived anthropomorphism

Anthropomorphism refers to the action of assigning or attributing human-like features or traits to entities that are not human, such as animals, objects, or machines [11]. This perception can be shaped by implicit social cognition which involves automatic and unconscious processes that support everyday social interactions [20]. It goes beyond the physical resemblance and encompasses the idea that people perceive non-human entities as having a theory of mind that refers to the capacity to understand and predict the behavior of others by attributing them to mental states like beliefs, desires, and intentions [37].

Perceived anthropomorphism can be categorized into two types: mindful and mindless [25]. Mindful anthropomorphism refers to intentionally attributing human-like characteristics to non-human objects. It involves deliberate thinking processes and recognizing that the person is applying social rules to something that is not inherently human [30]. On the other hand, mindless anthropomorphism refers to the unconscious attribution of human-like characteristics to non-human entities [25]. It can be triggered unconsciously by simple cues or heuristics [41]. In the context of human-computer interaction, perceived anthropomorphism has been shown to yield various benefits to the interaction. For instance, anthropomorphic website design (e.g., having human-like eyes on the web page) increased purchase intention by fostering trust through the attribution of anthropomorphic mental states of the website [46]. In another study, the use of anthropomorphic language (i.e., conversational versus impersonal) promoted information disclosure [39].

Previous research showed the importance of chatbot personification features, such as name or background information. For instance, [2] showed that assigning a chatbot a human name and human-like dialogical cues led to higher levels of perceived mindful and mindless anthropomorphism that, in turn, led to more favorable attitudes toward the chatbot and the company that employed it. Similarly, [7] demonstrated that a robot assigned with a name and background information evoked higher empathic concern toward itself. Based on the reviewed literature, we propose our first hypothesis:

H1: Chatbot personification leads to higher perceived a) mindful and b) mindless anthropomorphism.

## 2.2 Mediation effect of perceived anthropomorphism on willingness to donate

The willingness to donate (WTD) extends to the degree of inclination and emotional attachment an individual has toward offering assistance and contributing financially to a recipient who is gathering funds for the benefit of others [21].WTD is a complex construct that is influenced not only by individual factors, such as an individual's cognitive ability or their emotional state and personal relationship with the non-profit organization [34], but also by the traits from the agents – either human [10] or AI agent [3], [35]. In the case of AI agents, previous research has shown that AI interfaces containing anthropomorphic cues can lead to greater WTD compared to those that do not. For instance, [3] showed that people felt more psychological closeness with and were more willing to donate to AI agents with human-like appearances compared to the ones with machine-like appearances. In another study, participants who talked to a chatbot with a name perceived the chatbot as more human-like, increasing the likelihood of donating [40].

Other studies suggest that people mindlessly apply social heuristics to machines, for example, exhibiting politeness and reciprocity toward computers [30]. Further research in human-robot interaction demonstrates that people modify their behavior based on social pressure [43]. [19], for instance, showed that the perceived anthropomorphism of a robot would lead to higher conformity, i.e., a more salient social response to a non-human entity. In the context of donation, the presence of an observer was shown to modify people's altruistic intentions, specifically that people would donate more under social pressure [8]. Based on these findings, we suggest that a chatbot that has a name and background information will be perceived as anthropomorphic which, in turn, will lead to a higher WTD. Moreover, we posit that this effect will hold for both anthropomorphism types: When participants intentionally attribute human-like characteristics to the chatbot (mindful anthropomorphism) and unconsciously attribute human-like characteristics (mindless anthropomorphism). Thus, our second hypothesis is as follows:

H2: Perceived a) mindful and b) mindless anthropomorphism will mediate the effect of chatbot personification on WTD.

## 2.3 Mediation effect of chatbot perceptions on willingness to donate

Previous research suggests that chatbot perceived characteristics influence people's attitudes toward an AI agent. For instance, the perceived competence of a chatbot leads to less skepticism toward the agent, which in turn leads to more trust in the chatbot [36]. Another study showed that users experience a higher level of enjoyment when they perceive a chatbot as warm during the interaction [16].

Moreover, the research has shown that chatbots designed with anthropomorphic cues lead to more a positive human-chatbot interaction. For instance, [6] demonstrated that the chatbot's human-like interaction style increased the users' chatbot perceptions of trust and satisfaction. Another study found that a chatbot with a human name evoked a greater impression of it as competent, confident, sincere, and warm [40]. [40] demonstrated greater WTD to a chatbot with a human name, however, researchers did not investigate the mediation role of chatbot perceptions on the WTD. In order to fill this gap, in this study, we suggest that chatbot personification will lead to more positive chatbot perceptions that, in turn, will mediate subjects' WTD. Thus, we propose the following hypotheses:

H3: Chatbot personification leads to more favorable chatbot perceptions.

H4: Chatbot perceptions will mediate the effect of chatbot personification on WTD.

## 2.4 Effectiveness of different interpersonal persuasion strategies with chatbots

In the context of human-human communication, [45] defines interpersonal persuasion as the process in which two or more people communicate with the intention of changing the attitudes and/or behaviors of others. Interpersonal persuasion has been more effective compared to traditional mass media persuasion as it provides immediate feedback and coherence of behaviors. Today, various industries utilize chatbots to interpersonally transmit their persuasive messages. In turn, research has examined how different traditional persuasion strategies may differentially apply in human-chatbot scenarios.

Drawing upon previous work highlighting the importance of consistency in computer-agent interactions, the relative effectiveness of a given persuasion strategy may depend upon the relative personification of the chatbot. Classic work in social psychology suggests that inconsistency in traits requires more processing effort that, in turn, leads to unfavorable attitudes and negative behavioral responses [12]. Such effects appear to underlie user interactions with computer agents as well. For instance, inconsistent cues between agents' posture and verbal message can lessen an agent's persuasiveness and lead to more negative attitudes toward it [22]. Further, research examining human and synthetic voice and face pairings showed that an inconsistent agent (e.g., an agent with either a synthetic voice and human face or a human face and synthetic voice) was perceived as less trustworthy and stranger than those with congruent social cues [15]. Similarly to cues incongruity, inconsistency in tasks and agents can also lead to negative attitudes. For instance, AI agents involved in tasks requiring "human" capabilities, such as intuition, affect, or empathy, are perceived as less functionally competent [5]. On the other hand, AI agents that are utilized for tasks that are more objective in nature are perceived as more credible and freer from bias [41].

The current study investigates whether these findings related to consistency hold for a personified or non-personified chatbot agent that uses either a logical (objective task) or emotional (subjective task) persuasion strategy. Based on the previous findings, we propose to explore whether the persuasion strategy will have an interaction effect with chatbot personification. Thus, we propose the following questions:

RQ1: Will persuasion strategy moderate the influence of chatbot personification on the perceived a) mindful and b) mindless anthropomorphism of the chatbot?

RQ2: Will persuasion strategy moderate the influence of chatbot personification on chatbot perceptions?

## 3 METHOD

### 3.1 Design & Procedure

We conducted a 2 (personified vs. non-personified chatbot) x 2 (emotional vs. logical persuasion strategy) factorial design between-subjects laboratory experiment. Participants were first greeted in a reception area in which they first read and signed an informed consent document approved by the university's Institutional Review Board. Participants were then escorted to a dedicated study space and seated at a laptop workstation station. They then read through instructions outlining the study procedure. Once the experiment started, the participants were randomly assigned to one of four conditions and shown a mock-up website of an existing non-profit organization aiming to rescue homeless animals. Participants were prompted to start a conversation with the website's chatbot, operated via a Wizard-of-Oz design [33] in which the chatbot's responses were typed by a researcher who was present in a different room. These responses strictly followed a predetermined script and dialogue flow. After the conversation ended, the participants were asked to complete a brief questionnaire regarding their experience with the chatbot, trait measures, and demographics.

### 3.2 Participants

For this study, we recruited 86 participants from a college research participant pool at a private university in the northeastern United States. As compensation for their participation, individuals were granted one course credit. The final sample consisted of 76 participants after deleting those who did not pass the manipulation check. The majority of our sample participants self-identified as Asian ($N = 45$, 59.21%); 18 (23.68%) of participants were White; 8 (10.53%) were Hispanic; and 5 (6.58%) selected others or preferred not to report. Participants' age ranged from 18 to 29 years old ($M = 21,06$; $SD = 2.39$). In addition, 80.26% ($N = 61$) were female, 15.79% ($N = 12$) male, 2.63% ($N = 2$) non-binary/third gender, and 1.32% ($N = 1$) preferred not to report.

### 3.3 Manipulation of Chatbot Personification

Aligning with previous studies, this study manipulated name and background information of the chatbot agent [7], [2]. This manipulation was chosen based on its demonstrated effectiveness in inducing perceived anthropomorphism (both mindful and mindless) of the chatbot [2]. We designed two different conditions to examine people's perceptions of chatbots. In the personified chatbot conditions, the chatbot interface included a robot profile avatar, greeted participants by introducing its name as Benji, and provided them with brief background information about itself (i.e., "I love animals and have been working for Animal Rescue League for two years. How are you today?"). In the non-personified chatbot conditions, the chatbot did not have an avatar and simply greeted the participant with "Hi! How are you today?". In the following survey after the study, we included questions about the name and avatar of the chatbot, such as "Did the chatbot you just interacted with have a name/avatar?" to check if stimuli are effective enough for participants to notice (see Figure 1).

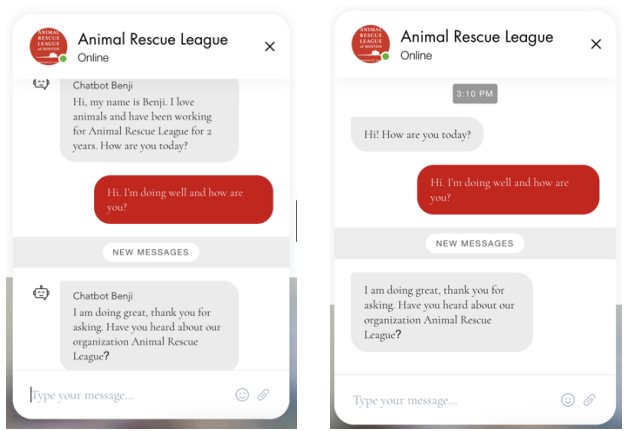

Figure 1: Chatbot Manipulation. On the left, an example of personified chatbot; on the right, an example of non-personified chatbot.

### 3.4 Manipulation of Persuasion Strategies

In order to manipulate persuasion strategies, the chatbot employed either a logical or an emotional persuasion appeal in their conversation with the participant. The script for chatbots using a logical strategy had the chatbot provide data and statistics about the current living situation of homeless animals, highlighting the number of animals entering shelters and the adoption rates (i.e., "Of the 6.3 million companion animals who enter shelters nationwide every year, only about 4.1 million animals are adopted"). The impact of the donation was also emphasized in numbers (i.e., "With the help of our donors, 20,653 animals were treated, adopted or provided with advocacy for their rights").

The script for chatbots employing the emotional strategy instead focused on sharing a moving story about rescuing an animal, emphasizing the challenging aspects of animal rescue work and the need for more help. In this condition, the chatbot provided the following additional details regarding animal rescue: "Once during the rescue, we had no choice but to cut off one of the legs of a cat because it was seriously infected. With the help of our donors, we also confront animal cruelty and neglect through special police investigation. We work hard but would appreciate more help."

### 3.5 Measures

*3.5.1 Willingness to donate.*

Before the start of interaction with the chatbot, participants were told that they would be entering a raffle for $10. During the conversation with the chatbot, they were asked how much of this money they would like to donate to Animal Rescue League. The donation amount varied from a minimum $0 (5.26% of participants) to $10 (81.58% of participants). The donation mean was 8.66, with a standard deviation of 3.01.

*3.5.2 Chatbot perceptions.*

We employed a combination of two existing self-report questionnaires of chatbot impressions [40] and chatbot experience [17] to construct a chatbot perceptions measure. Chatbot impressions were captured via 4-item 7-point semantic differential scale (incompetent–competent, unconfident–confident, cold–warm, and insincere–sincere; $M = 5.10$, $SD = 1.04$, $\alpha = .84$). Chatbot experience, we measured through 4-item 7-point Likert scale gauging the extent to which participants found the chat experience to be interesting, entertaining, enjoyable, and pleasant ($M = 4.52$, $SD = 1.29$, $\alpha = .93$). The chatbot impressions and chatbot experience measures were computed into an aggregated variable chatbot perceptions ($M$ = 4.81, $SD$ = .13, $α$ = .88).

*3.5.3 Perceived anthropomorphism.*

To measure perceived anthropomorphism, we adopted two measures, one of mindful anthropomorphism and one of mindless anthropomorphism [25]. The two measures consist of a series of 7-point semantic differential scales. The measure for mindful anthropomorphism ($M$ = 3.49, $SD$ = 1.30, $α$ = .86) consists of multiple 7-point semantic differential scales by which the participants evaluated the chatbot in terms of the extent to which it was humanlike–machinelike, natural–unnatural, lifelike–artificial. The mindless anthropomorphism measure ($M$ = 4.14, $SD$ = 1.08, $α$ = .89) consisted of a series of 7-point Likert style items asking participants to evaluate the extent to which the chatbot could be described as attractive, exciting, pleasant, interesting, likable, sociable, friendly, and personal.

*3.5.4 Control variables.*

Control variables play a pivotal role in research, as they account for extraneous factors that may influence study outcomes. Previous studies have found that higher empathic concern levels lead to increased donations [42], and individuals with higher empathic concern are more strongly connected to AI agents with background information [7]. Moreover, prior experience has been shown to impact people's evaluations of chatbots. For instance, [23] discovered that after controlling for familiarity, significant differences emerged in perceived anthropomorphism between chatbots and websites, such that participants who were less familiar with chatbot technology assigned lower scores to chatbots. In turn, to account for the effect of such individual difference variables, the current study controlled for the influence of participants' level of trait empathy as a covariate for donation outcome and participants' familiarity with chatbots as a covariate for chatbot perceptions.

*3.5.5 Trait Empathy.*

To assess trait empathy, we utilized the Interpersonal Reactivity Index subscales, empathetic concern, and personal distress [42]. Participants rated their agreement with a series of statements, with higher scores indicating greater empathy ($M$ = 5.22, $SD$ = .89, $α$ = .79), including survey questions such as concern for the less fortunate, emotional responses to others' problems, protective instincts, and self-perception as a soft-hearted person.

*3.5.6 Familiarity.*

To measure familiarity with chatbots, we adapted a 7-point scale from [48]. Participants rated their level of agreement with statements concerning their experience and understanding of chatbots ($M$ = 5.50, $SD$ = .70, $α$ = .69), and answered questions about their understanding of chatbot functionality, previous interactions with chatbots, ability to differentiate between chatbots and humans, trust in chatbot-generated information, and awareness of chatbots' limitations.

## 4 RESULTS

This study utilized the PROCESS Model 7 [18] to examine the relationship between the predictor variable–personification, and the outcome variable–willingness to donate (WTD), while considering the mediators–chatbot perception and anthropomorphism (mindful and mindless). The persuasion strategy is considered a moderating variable between personification and the mediators. The covariates familiarity and empathy were also included in the model. The analyses were conducted at a 95% confidence level, and bootstrap sampling was used for percentile bootstrap confidence intervals with 5000 samples (see Figure 2).

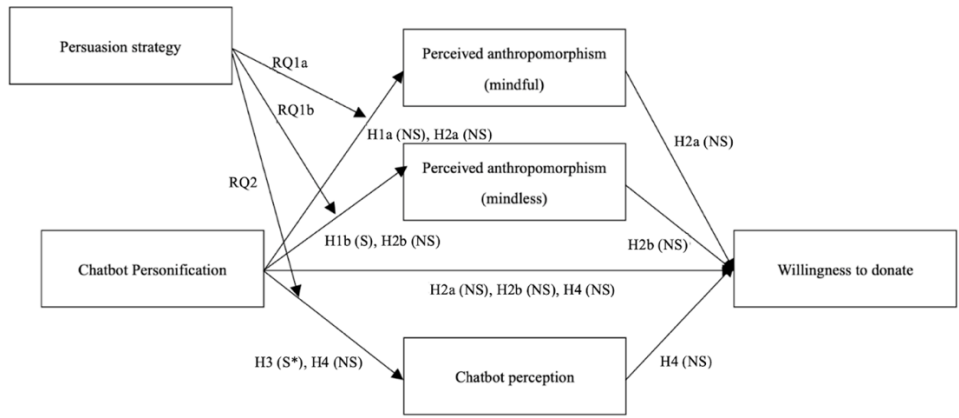

* NS = not significant; S = significant in predicted direction; S* = significant in the opposite direction

Figure 2: Causal diagram with tested hypotheses (see Figure Description in Appendix).

### 4.1 The effect of chatbot personification on perceived anthropomorphism

*4.1.1 Mindful Anthropomorphism.*

The model predicting mindful anthropomorphism was not significant, $F(5, 70) = 1.98$, $p = .09$, explaining 12.41% of the variance ($R^2 = .12$). Neither chatbot personification, persuasion strategy, nor the interaction between personification and persuasion was a significant predictor of mindful anthropomorphism when controlling for familiarity and trait empathy. Thus, chatbot personification was not found to enhance mindful anthropomorphism (H1a), nor was such an effect moderated by persuasion strategy (RQ1a).

*4.1.2 Mindless Anthropomorphism.*

The model predicting mindless anthropomorphism was significant, $F(5, 70) = 3.38$, $p = .01$, explaining 19.45% of the variance ($R^2 = .19$). Chatbot personification, $\beta = -1.69$, $t(70) = -2.32$, $p = .02$, and persuasion strategy, $\beta = -1.45$, $t(70) = -1.99$, $p = .05$, were found to be significant predictors of mindless anthropomorphism, when controlling for familiarity, $\beta = .47$, $t(70) = 2.70$, $p = .01$ and trait empathy, $\beta = -.05$, $t(70) = -.39$, $p = .70$ Thus, H1b was supported. The interaction effect between chatbot personification and persuasion strategy was also significant, $\beta = 1.15$, $t(70) = 2.42$, $p = .02$. However, though the combined effect of personification and persuasion strategy was significant, pairwise comparisons revealed that the effect of chatbot personification was not independently significant within the logical strategy ($p = .10$) or emotional strategy conditions ($p = .09$). This suggests a complex nature of their interaction between personification and persuasion strategy in influencing mindless anthropomorphism of chatbots (RQ1b).

*4.1.3 Chatbot Perception.*

The model predicting chatbot perceptions was significant, $F(5, 70) = 7.45$, $p < .001$, explaining 34.72% of the variance ($R^2 = .34$). Chatbot personification, $\beta = -1.36$, $t(70) = -2.27$, $p = .03$, was a significant predictor of chatbot perceptions, but persuasion strategy was not, $\beta = -1.02$, $t(70) = -1.70$, $p = .09$, when controlling for familiarity, $\beta = .65$, $t(70) = 4.56$, $p < .001$ and trait empathy ($\beta = .15$, $t(70) = 1.44$, $p = .16$). The result showed that chatbot personification actually led to less favorable chatbot perceptions, counter to H3. With respect to RQ2, the interaction effect between chatbot personification and persuasion strategy was significant, $\beta = .979$, $t(70) = 2.01$, $p < .05$, such that the negative effect of personification on chatbot perceptions was observed when the chatbot employed a logical persuasive strategy ($p = .03$) but not when leveraging an emotional strategy ($p = .48$).

## 4.2 The effects of personification and persuasion strategy on WTD through chatbot perceptions and anthropomorphism

Mediation analysis was also completed to examine the causal pathways described above. The direct effect of chatbot personification on WTD was again not significant, $R^2 = .27$, $t(69) = .39$, $p = .70$. The indirect effects of chatbot personification on WTD through mindful anthropomorphism (logic: BootLLCI = -.44, BootULCI = .37; emotional: BootLLCI = -.35, BootULCI = .72), mindless anthropomorphism (logic: BootLLCI = -1.44, BootULCI = .27; emotional: BootLLCI = -.25, BootULCI = 1.83), and chatbot perceptions (logic: BootLLCI = -1.08, BootULCI = .42; emotional: BootLLCI = -.26, BootULCI = .71) were not statistically significant at the 95% confidence level.

## 5 DISCUSSION

The present study investigates how chatbot personification, different persuasion strategies, and chatbot perceptions influence people's WTD and what are causal relationships among these variables. The results suggest that interaction with a chatbot with a name and background information does not elicit higher WTD or more favorable chatbot perceptions. In fact, our findings demonstrate that commonly used anthropomorphic features, such as name and narrative, can lead to negative attitudes toward an AI agent. These results align with previous empirical literature that found a negative effect of anthropomorphic cues in human-machine interaction settings [32],[4].

Further, our study highlights the role of perceived anthropomorphism in the interaction with a chatbot agent. The mediation analysis revealed that chatbot personification evokes mindless, but not mindful, anthropomorphism, however, perceived mindless anthropomorphism does not lead to WTD. Thus, these findings provide evidence for theory-driven relationships between anthropomorphic cues and perceived anthropomorphism [25] and previous suggestion that users tend to treat computers as human beings non-consciously rather consciously [29]. However, in the donation context, the chatbot that was mindlessly perceived as anthropomorphic did not lead to a desirable outcome. This highlights the importance of the context in which chatbots are utilized and the final objectives of the interaction. For instance, a study placed in the customer service setting showed that a chatbot with anthropomorphic cues had a positive effect on customers attitudes e.g., [2], while a chatbot with anthropomorphic language cues diminished the people's intent to donate in the context of charity [49].

Further, our findings demonstrate that persuasion strategies moderate the influence of chatbot personification on the perceived mindless, but not mindful anthropomorphism of the chatbot. Notably, our analysis revealed an interaction between chatbot personification and persuasion strategy. While individual pairwise comparisons were non-significant, the non-personified chatbot paired with a logical persuasion strategy led to the most favorable chatbot perceptions. This result resonates with [15]'s argument that the consistency between visual and verbal cues is important for producing favorable attitudes regarding agent-based user interfaces. Moreover, this result supports the literature on machine heuristics that suggests people hold perceptions of machines as more accurate and objective compared to humans, and thus, more suitable for the logical rather than emotional appeals [28]. Thus, it is conceivable that due to machine heuristics, the non-personified chatbot paired with the logical persuasion strategy that involved statistical information was perceived more favorable compared to a personified chatbot.

Of course, this begs the question of why a comparable congruency effect was not observed when comparing personification levels across the emotional persuasion strategy conditions. This might be attributed to what we have mentioned above: The fact that personification cues employed in the experiment (name and background information) only facilitated the mindless perceived anthropomorphism of a chatbot. As chatbot personification had no direct effect on mindful anthropomorphism—that is, users are consciously aware of the non-humanness of the chatbot—it may be the case that a chatbot using an emotional strategy in requesting donations from users was perceived as inherently inconsistent with its nonhuman nature, regardless of personification. In other words, the congruency effect may not manifest when other elements of the interaction suggest a greater baseline inconsistency in the chatbot.

## 6 THEORETICAL AND PRACTICAL IMPLICATIONS

The results of this study have important theoretical implications for both persuasion and anthropomorphism. Our study highlights the possibility of the negative effect of anthropomorphic cues implementation. Further, our results demonstrate the importance of matching the persuasion strategy used in chatbot interactions with the context in which they are employed. Our findings provide further empirical support for theoretical accounts of the importance of congruency of cues in multifaceted interactions, not just for human-human exchanges but, as observed here, in human-machine communication contexts.

Further, this study supports the theoretical relationships between anthropomorphic cues and perceived anthropomorphism, emphasizing user tendency toward mindless rather than mindful anthropomorphism. However, in the context of the donation likelihood, the well-established anthropomorphic cues were not found to enhance desirable effects (in our case, WTD). As such, the current findings suggest that some of the most common anthropomorphic design features might need further refining if they are to effectively leverage these psychological mechanisms toward practical application. The results of this study also highlight additional practical implications for the use of chatbots in the context of charitable donations. Our study suggests exercising caution in implementation of human-like cues when constructing the image of chatbot agents and the dialogue flows. First, anthropomorphic cues might lead to negative chatbot perceptions. Second, our results demonstrate that a non-personified chatbot paired with the logical persuasion strategy may be more appropriate in the context of donation likelihood.

Finally, our study gives timely directions for designing chatbots that align with the current regulations of AI systems in the face of the potential risks they pose. With the advancement of large language models that can produce text indistinguishable from human writing, the current regulations oblige companies to disclose chatbots' identities [26]. Further, the research shows that excessive anthropomorphism of chatbots and users' encouragement to relate to the AI systems as human beings might lead to transparency and trust issues, and high risks of over-reliance on these systems [1]. While the majority of chatbot studies investigate effects of chatbot versus human interaction, our study gives important insights into chatbot design under conditions when a chatbot does not disguise as a human agent.

# 7 LIMITATIONS AND FUTURE RESEARCH

The current study has a number of important limitations that may be addressed in future work. First, our study sample is relatively small, homogeneous, and recruited through a single university research pool. As such, the sample may not be representative of typical visitors of NGO websites, potentially limiting generalizability to certain donation markets. Secondly, the majority of the current sample's participants were willing to donate the maximum amount of $10 (81.6% of participants), leading to a skewed distribution of the WTD variable. Follow-up studies should be conducted with a broader yet targeted sample in order to corroborate the current findings. Ultimately, our findings contribute to a deeper understanding of how theoretical accounts of persuasion and anthropomorphism can inform the design and implementation of chatbots in the domain of charity and donation. The present study provides a number of suggestions for the future development of more effective chatbots for nonprofit fundraising.

**APPENDIX**

**Figure 2 Description**

1. Persuasion strategy → Perceived anthropomorphism (mindful) (RQ1a)
2. Persuasion strategy → Perceived anthropomorphism (mindless) (RQ1b)
3. Persuasion strategy → Chatbot perception (RQ2)
4. Chatbot Personification → Perceived anthropomorphism (mindful) (H1a (NS), H2a (NS))
5. Chatbot Personification → Perceived anthropomorphism (mindless) (H1b (S), H2b (NS))
6. Chatbot Personification → Willingness to donate (H2a (NS), H2b (NS), H4 (NS))
7. Chatbot Personification → Chatbot perception (H3 (S*), H4 (NS))
8. Perceived anthropomorphism (mindful) → Willingness to donate (H2a (NS))
9. Perceived anthropomorphism (mindless) → Willingness to donate (H2b (NS))
10. Chatbot perception → Willingness to donate (H4 (NS))

NS = not significant; S = significant; S* = significant in the opposite direction